\documentclass[11pt]{article}

\usepackage{graphicx}
\usepackage[figuresright]{rotating}

\title{New Light on Dark Matter from the LHC}
\author{John Ellis}
\date{Theory Division, CERN, CH--1211 Gen\`eve 23, Switzerland; \\ 
Theoretical Physics and Cosmology Group, Department of Physics, KingÕs College London, London WC2R 2LS, UK \\ ~ \\
October 2010 \\ ~ \\
\begin{center}
{\tt CERN-PH-TH/2010-258~~~~~~~~~~~~~~~~~~~~KCL-PH-TH/2010-31}
\resizebox{1.5cm}{!}{\includegraphics{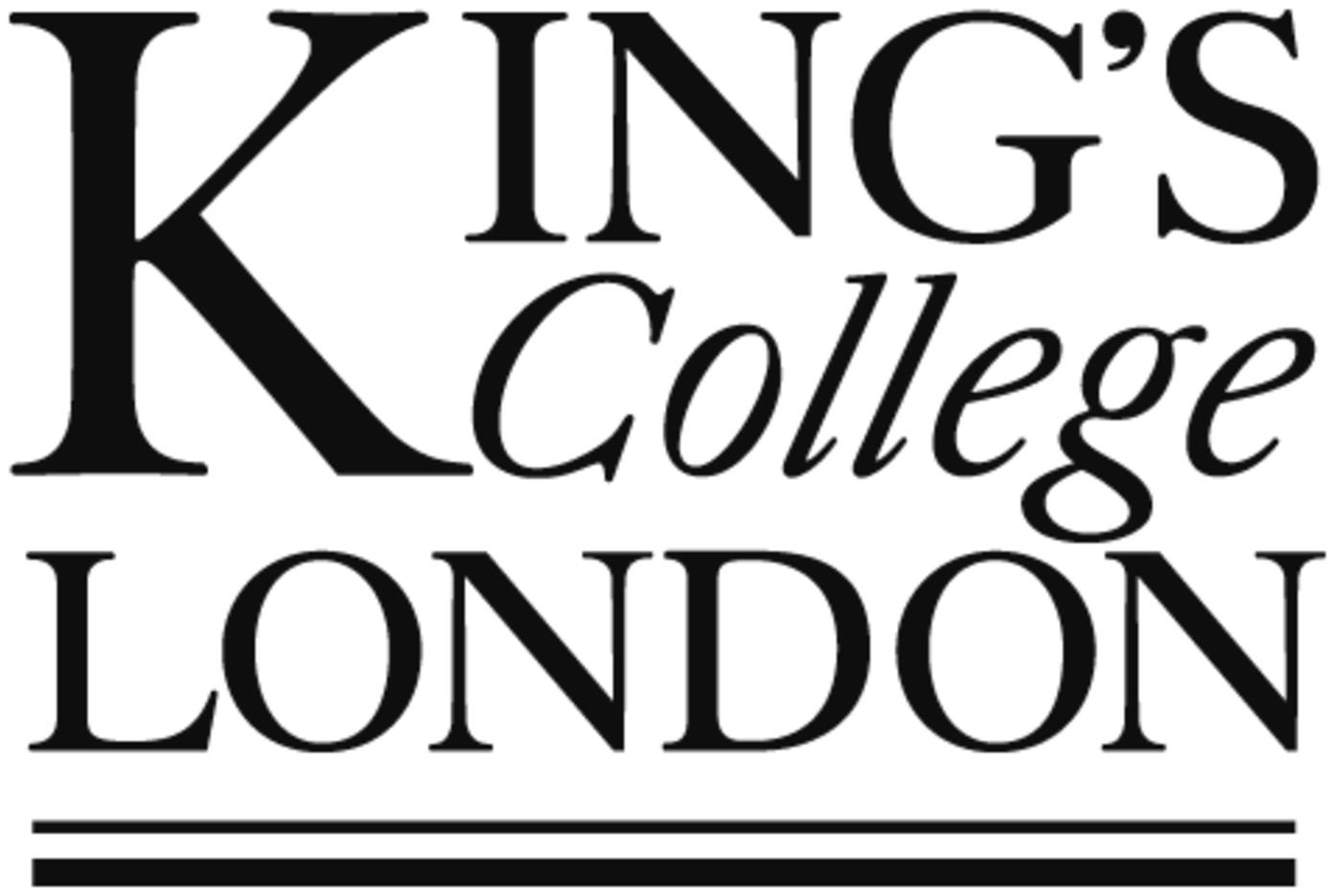}}~~~~~
\end{center}
\vspace{-2cm}
}

\begin{document}
\maketitle

\begin{center}
{\bf Abstract}
\end{center}

The prospects for detecting a candidate
supersymmetric dark matter particle at the LHC are reviewed, and compared
with the prospects for direct and indirect searches for astrophysical dark matter,
on the basis of a frequentist analysis of the preferred regions of the Minimal
supersymmetric extension of the Standard Model with universal soft
supersymmetry breaking (the CMSSM) and a model with equal but non-universal
supersymmetry-breaking contributions to the Higgs masses (the NUHM1).
LHC searches may have good chances to observe supersymmetry in the near future
- and so may direct searches for astrophysical dark matter particles.

\section{Introduction}\label{s:intro}

There is a standard list of open questions beyond the Standard Model
of particle physics~\cite{COlex}, which includes the following.
(1) What is the origin of particle masses and, in particular, are they due to a Higgs boson?
(2) Why are there so many different types of standard matter particles, notably three neutrino species?
(3) What is the dark matter in the Universe?
(4) How can we unify the fundamental forces?
(5) Last but certainly not least, how may we construct a quantum theory of gravity?
Each of these questions will be addressed, in some way, by experiments at the LHC,
though answers to all of them are not guaranteed!

The central topic of this talk is, of course, question (3) concerning dark matter.
Certainly there are many candidate particles, ranging in mass from axions to
Wimpzillas. However, many candidates fall within the general category of WIMPs 
(weakly-interacting massive particles) weighing between $\sim 100$ and
$\sim 1000$~GeV and hence possibly accessible to the LHC. These include
the lightest Kaluza-Klein particle (LKP)
in some scenarios with extra dimensions~\cite{LKP}, the lightest T-odd particle (LTP)
in some little Higgs scenarios~\cite{LTP}, and the lightest supersymmetric particle (LSP)
in supersymmetric models in which R-parity is conserved~\cite{LSP}.

Historically, the LSP was the first of these WIMP candidates, and personally
I still find the LSP the best motivated,
since there are so many reasons to favour supersymmetry at the TeV scale~\cite{COlex}.
It would help the Higgs boson do its job [(1) above], by cancelling the quadratically-divergent
contributions to its mass, and thereby stabilizing the electroweak mass scale~\cite{hierarchy}.
Further, supersymmetry predicts the appearance of a Higgs boson at a mass
$\sim 130$~GeV or below, as indicated by the precision electroweak data~\cite{lightH}.
Supersymmetry at the TeV scale would also aid in the grand unification of the
strong, weak and electromagnetic interactions~\cite{GUTs} by enabling their strengths to
evolve to a common value at some high-energy GUT scale [(4) above]. Moreover, supersymmetry is
apparently essential in stringy attempts to construct a quantum theory of
gravity  [(5) above]. However, as Feynman surely would have said, you would not give
five arguments for supersymmetry if you had one good argument, so let us focus on that: the
LSP is an excellent candidate for dark matter  [(3) above]~\cite{LSP}, as we now discuss.

\section{Supersymmetric Models}\label{s:models}

We work within the framework of the minimal supersymetric extension
of the Standard Model (MSSM), in which the known particles are
accompanied by simple supersymmetric partners and there are two
Higgs doublets, with a superpotential coupling denoted by $\mu$ and
a ratio of Higgs v.e.v.s denoted by $\tan \beta$~\cite{MSSM}. The bugbear of the MSSM
is supersymmetry breaking, which occurs generically through scalar
masses $m_0$, gaugino fermion masses $m_{1/2}$, trilinear soft
scalar couplings $A_0$ and bilinear soft scalar couplings $B_0$.
In our ignorance about them, the total number of parameters in the MSSM 
exceeds 100! For simplicity,
it is often assumed that these parameters are universal at the scale of
grand unification, so that there are single soft supersymmetry-breaking
parameters $m_0, m_{1/2}, A_0$, a scenario called the constrained
MSSM (CMSSM)~\footnote{I emphasize that the CMSSM is
not to be confused with minimal supergravity (mSUGRA), which
imposes a specific relationship between the trilinear and bilinear
couplings: $B_0 = A_0 - m_0$ as well as a relationship between
the scalar and gravitino masses: $m_0 = m_{3/2}$. These apparently
innocuous extra assumptions affect drastically the nature of the LSP,
and the allowed regions of parameter space~\cite{VCMSSM}.}. 
However, this assumption is not strongly motivated
by either fundamental theory or phenomenology. Moreover, as discussed below,
even if $m_0, m_{1/2}, A_0$ are universal, this may be true at some
scale different from the GUT scale~\cite{EOS2,EMO}.

What happens if the soft supersymmetry-breaking parameters are
not universal? Upper limits on flavour-changing neutral interactions
disfavour models in which different sfermions with the same internal
quantum numbers, e.g., the $ {\tilde d}, {\tilde s}$ squarks have different masses~\cite{EN}.
But what about squarks with different internal quantum numbers, or
squarks and sleptons? Various GUT models impose some relations between them,
e.g., the ${\tilde d_R}$ and ${\tilde e_L}$ scalar masses are universal in SU(5) GUTs, 
as are the ${\tilde d_L}, {\tilde u_L}, {\tilde u_R}$ and ${\tilde e_R}$ scalar masses,
and all are equal in SO(10) GUTs. However, none of these arguments rules out
non-universal supersymmetry-breaking scalar masses for the Higgs multiplets,
so one may also consider such non-universal Higgs models (NUHM) with either
one or two additional parameters (NUHM1, NUHM2). Who knows where
string models may finish up among or beyond these possibilities?

The LSP is stable in many supersymmetric models because of a conserved
quantity known as $R$ parity, which may be expressed in terms of baryon
number $B$, lepton number $L$ and spin $S$ as $R \equiv (-1)^{2S - L + 3B}$. 
It is easy to check that all Standard Model particles have $R = +1$ and their
supersymmetric partners have $R = - 1$. The multiplicative conservation of $R$
implies that sparticles must be produced in pairs that heavier sparticles must
decay into lighter sparticles, and that the LSP is stable, because it has no
legal decay mode. It should lie around in the Universe today, as a
supersymmetric relic from the Big Bang~\cite{LSP}.

In such a scenario, the LSP could have no strong or electromagnetic interactions~\cite{LSP},
since otherwise it would bind to ordinary matter and be detectable in 
anomalous heavy nuclei, which have been looked for, but not seen.
Possible weakly-interacting scandidates include {\it a priori} the sneutrinos -
which have been excluded by LEP and by direct astrophysical searches for
dark matter, the lightest neutralino $\chi$ - a mixture of the spartners of the
$Z, \gamma$ and neutral Higgs boson, and the gravitino - the supersymmetric
partner of the graviton, which would be a nightmare for astrophysical detection,
but a potential bonanza for collider experiments. Here we concentrate on the neutralino
option, whose classical signature is en event with missing transverse momentum
carried away by invisible dark matter particles. This signature is shared by other
WIMP candidates for dark matter, such as the LKP~\cite{LKP} and LTP~\cite{LTP}, though the nature
and kinematics of the visible stuff accompanying the dark matter particles is
model-dependent.

\section{Constraining Supersymmetry}\label{s:constraints}

There are significant lower limits on the possible masses of supersymmetric
particles from LEP, which requires any charged sparticle to weigh more than
about 100~GeV~\cite{LEPsusy}, and the Tevatron collider, which has not found any squarks
or gluinos lighter than about 400~GeV~\cite{Tevatron}. There are also important indirect
constraints implied by the LEP lower limit on the Higgs mass of 114.4~GeV~\cite{LEPH}, and the
agreement of the Standard Model prediction for $b \to s \gamma$ decay
with experimental measurements. The only possible experimental discrepancy
with a Standard Model prediction is for $g_\mu - 2$~\cite{E821}, though the significance
of this discrepancy is still uncertain, as discussed in the following paragraph. 
However, there is one clear discrepancy with the Standard Model of particles, namely the density
of dark matter, which cannot be explained without physics beyond the Standard
Model, such as supersymmetry. The fact that the dark matter density is
constrained to within a range of a few percent~\cite{WMAP}:
\begin{equation}
\Omega_{DM} \; = \; 0.111 \pm 0.006
\label{WMAP}
\end{equation}
constrains some combination of the parameters of any dark matter model also to within a few percent,
as we shall see shortly in the case of supersymmetry, but the same would be
true in other models.

The calculation of the Standard Model prediction for $g_\mu - 2$ requires an estimate of
the contribution from hadronic vacuum polarization diagrams, that may be obtained either
from $e^+ e^- \to$~hadrons data, or from $\tau \to \nu +$~hadrons decays. Historically,
there has been poor consistency between the $e^+ e^-$ and $\tau$ estimates (though
both differ substantially from the experimental measurement), and
the consistency between different $e^+ e^-$ experiments has not always been
excellent. Since the $\tau$ estimate requires an isospin correction, the $e^+ e^-$ 
estimate is more direct and generally preferred. Accordingly, in the following results are
shown assuming a discrepancy~\cite{oldDavier}
\begin{equation}
\Delta ( g_\mu - 2 ) \; = \; (30.2 \pm 8.8) \times 10^{-10}
\label{Deltag-2}
\end{equation}
calculated from $e^+ e^-$ data
to be explained by physics beyond the Standard Model, such as supersymmetry.
Very recently, re-evaluations of the $e^+ e^-$ and $\tau$ data
have yielded $\Delta ( g_\mu - 2 ) = (28.7 \pm 8.0) \times 10^{-10}$ 
and $(19.5 \pm 8.3) \times 10^{-10}$~\cite{newDavier},
corresponding to discrepancies of 3.6 and 2.4 $\sigma$, respectively. The
results shown below would differ very little if the newer $e^+ e^-$ estimate were used.
For comparison, some results from dropping the $g_\mu - 2$ constraint altogether
are also shown, and using the $\tau$ decay estimate would give intermediate
results closer to the $e^+ e^-$ estimate.

\begin{figure}
  \includegraphics[width=.45\textwidth]{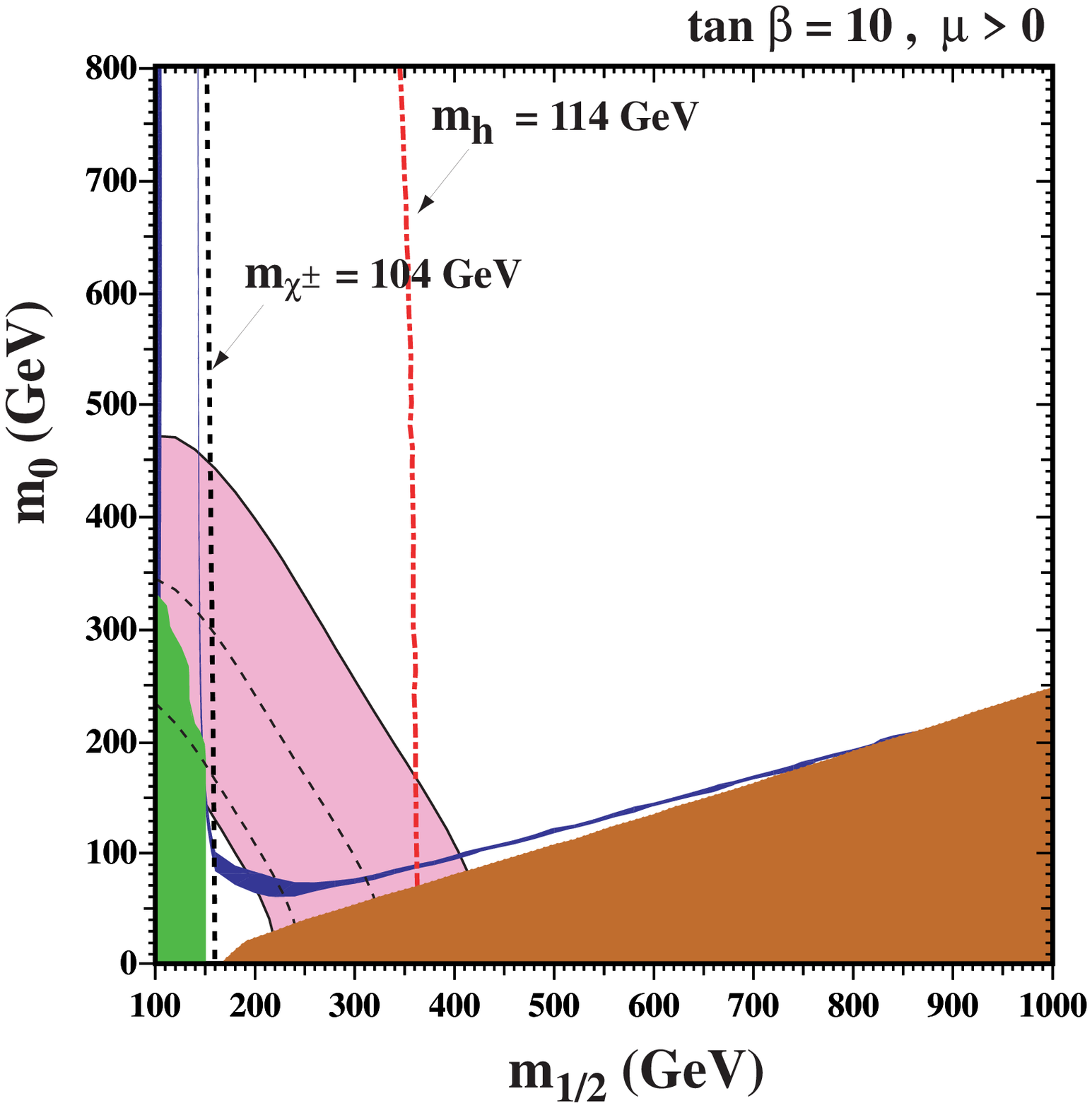}
    \includegraphics[width=.45\textwidth]{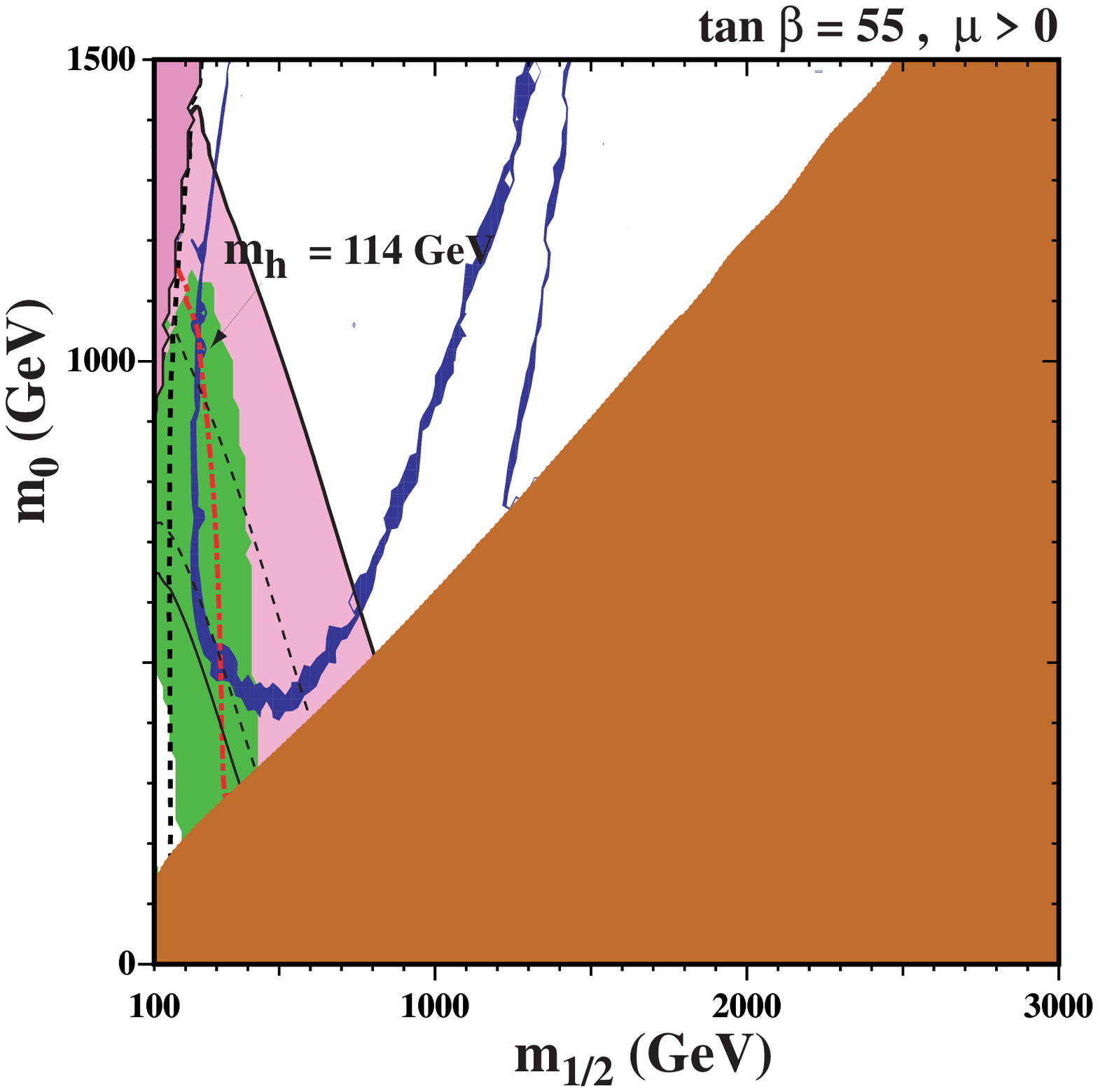}\\
      \includegraphics[width=.45\textwidth]{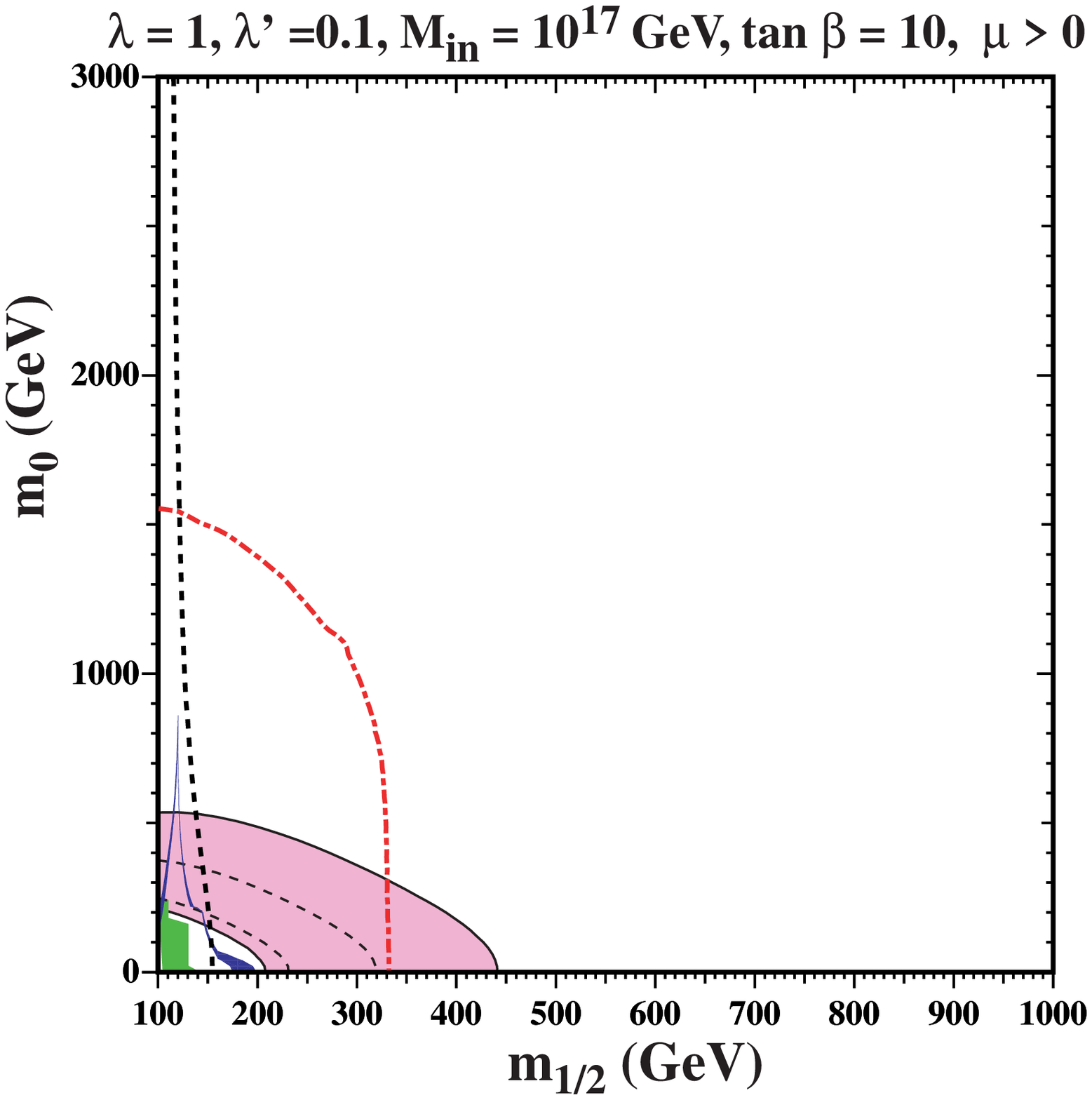}
            \includegraphics[width=.45\textwidth]{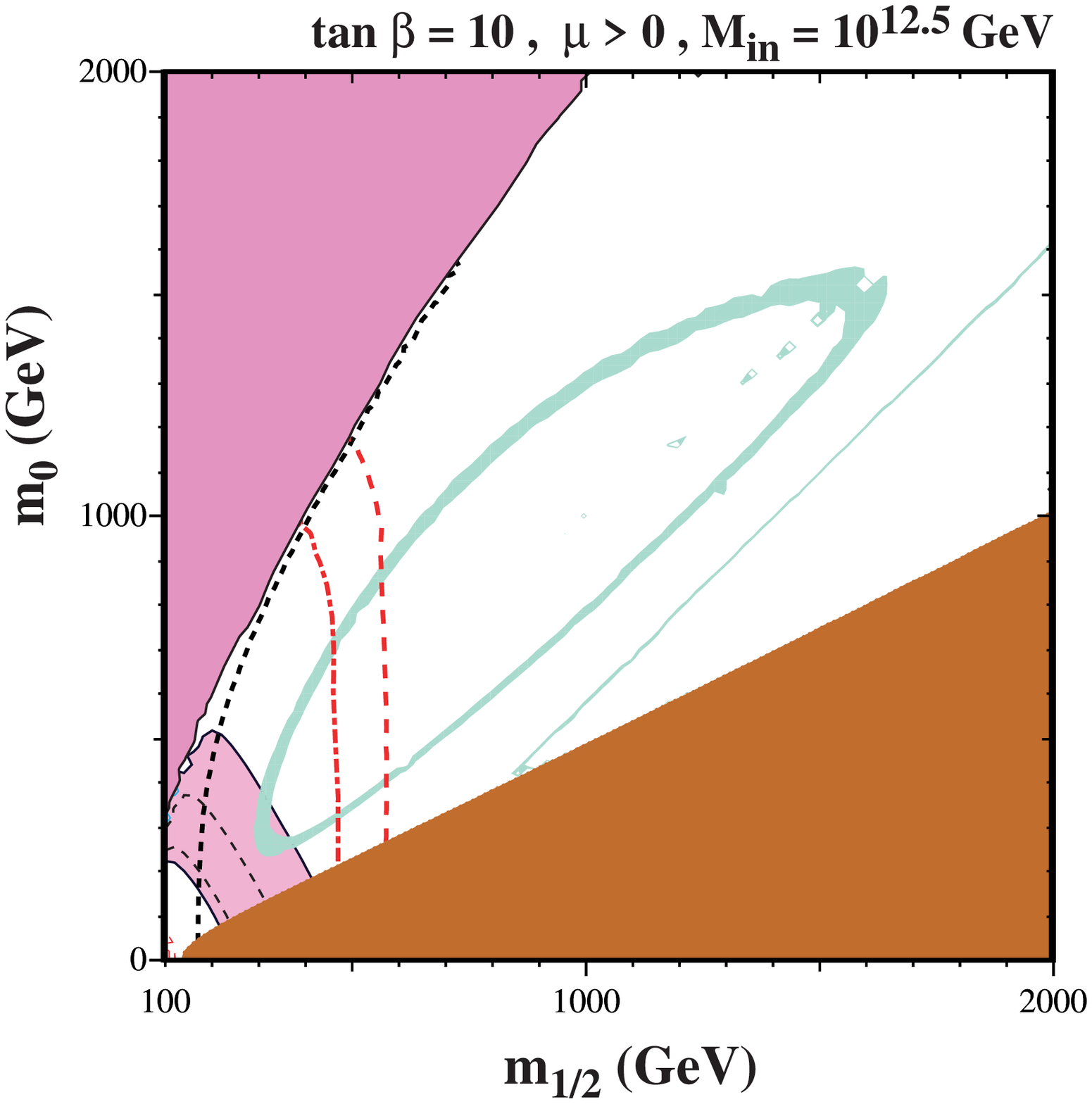}
  \caption{\it The $(m_{1/2}, m_0)$ planes for  (upper left) the CMSSM with $\tan \beta = 10$ and (upper right)
  $\tan \beta = 55$~\protect\cite{EOSS}, (lower left) assuming SU(5) universality at 
  $10^{17}$~GeV with representative
  choices of the quartic GUT Higgs couplings~\protect\cite{EMO}, 
  and (lower right) assuming scalar mass
  universality at $10^{12.5}$~GeV~\protect\cite{EOS2}, 
  all assuming  $\mu > 0, A_0 = 0, m_t = 173.1$~GeV and
$m_b(m_b)^{\overline {MS}}_{SM} = 4.25$~GeV. The near-vertical (red)
dot-dashed lines are the contours $m_h = 114$~GeV~\protect\cite{LEPH}, 
and the near-vertical (black) dashed
line is the contour $m_{\chi^\pm} = 104$~GeV~\protect\cite{LEPsusy}.  The medium (dark
green) shaded region is excluded by $b \to s
\gamma$, and the dark (blue) shaded area is the cosmologically
preferred region~\protect\cite{WMAP}. In the dark
(brick red) shaded region, the LSP is the charged lighter stau slepton. The
region allowed by the E821 measurement of $g_\mu -2$ at the 2-$\sigma$
level, is shaded (pink) and bounded by solid black lines, with dashed
lines indicating the 1-$\sigma$ ranges.}
  \label{fig:CMSSM}
  
\end{figure}

Fig.~\ref{fig:CMSSM} demonstrates the impacts of the various theoretical,
phenomenological, experimental and cosmological constraints in $(m_{1/2}, m_0)$
planes under different scenarios with $\mu >0$, assuming that the LSP is the
lightest neutralino, $\chi$. The top panels are for the CMSSM 
with $A_0 = 0$ and (left) $\tan \beta = 10$, (right) $\tan \beta = 55$, two values
that bracket the plausible range~\cite{EOSS}. In both cases, we see narrow WMAP-compliant
strips clinging near the boundaries of the (brown) charged LSP region at low $m_0$, 
where LSP-slepton coannihilation is important, and the (pink) region at high $m_{1/2}$
where electroweak symmetry is not broken consistently, called the focus-point strip.
When $\tan \beta = 55$, we also see a diagonal funnel at large $m_{1/2}$ and $m_0$
due to rapid annihilation through direct-channel heavy Higgs poles. In the lower
left panel, also for $\tan \beta = 10$, it is assumed that the scalar masses $m_0$
and the gaugino masses $m_{1/2}$ are universal at the scale $10^{17}$~GeV~\cite{EMO},
instead of the GUT scale as in the CMSSM. We see that the coannihilation strip
has shrunk into the region forbidden by the LEP Higgs limit, and the fixed-point
strip has disappeared to larger $m_0$. On the other hand, if $m_0$ universality
is assumed instead to hold at $10^{12.5}$~GeV, as in the bottom right panel, the
coannihilation, fixed-point and funnel regions merge to form an atoll away from
the boundaries of parameter space~\cite{EOS2}. In what follows, the standard CMSSM and
the NUHM1 model will be studied, but these panels emphasize that this involves
a dicey assumption.

\section{Global Supersymmetric Fits}

Within the general CMSSM and NUHM frameworks, is it possible to find
a preferred region of supersymmetric masses? To answer this question,
we adopted a frequentist approach and constructed a global likelihood function
using precision electroweak data, the LEP Higgs mass limit (allowing for 
theoretical uncertainties), the cold dark matter density, $b \to s \gamma$
and $B_s \to \mu^+ \mu^-$ constraints and (optionally) $g_\mu - 2$~\cite{MC1,MC2,MC3}.

\begin{figure*}[htb!]
\begin{center}
\vspace{-10cm}
\begin{picture}(300,400)
  \put(  -15,   0){ \resizebox{5.75cm}{!}{\includegraphics{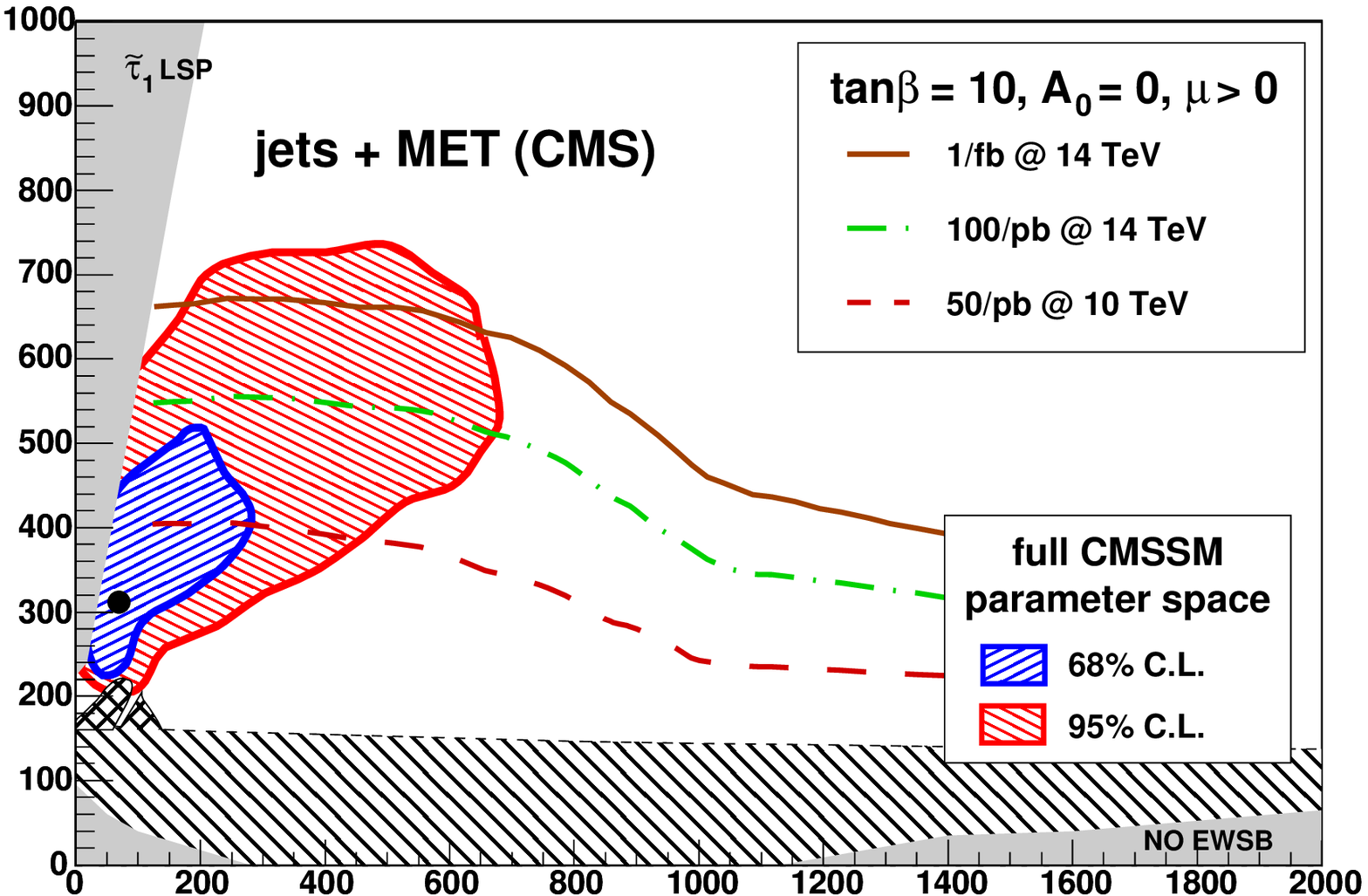}}}
  \put(100,   -12){$m_0$ [GeV]}
  \put( -20,   50){\begin{rotate}{90}$m_{1/2}$ [GeV]\end{rotate}}
  \put(  160,     0){ \resizebox{5.75cm}{!}{\includegraphics{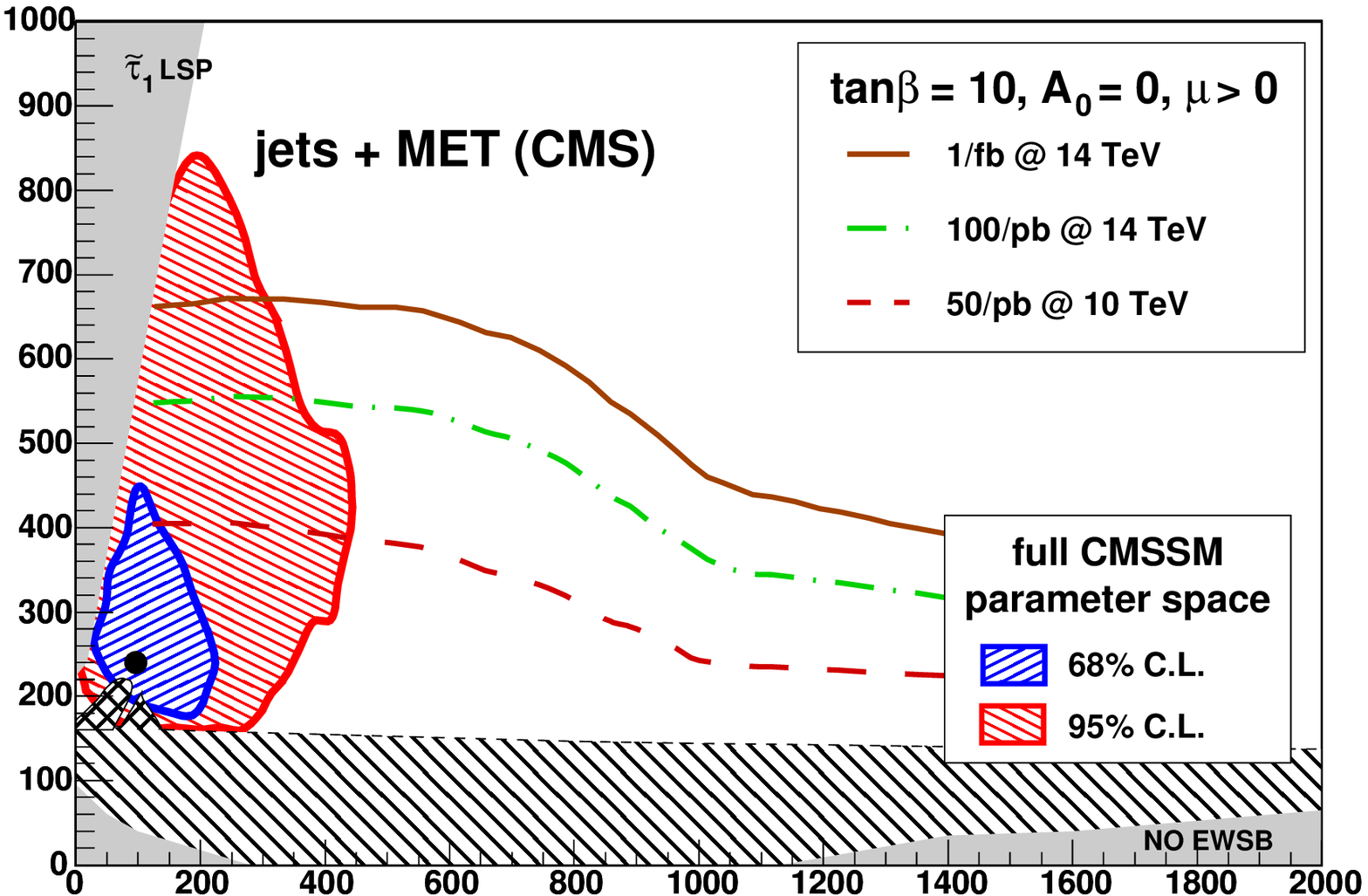}}  }
  \put(280,   -12){$m_0$ [GeV]}
\end{picture}
\end{center}
\caption {\it The $(m_0, m_{1/2})$ planes for (left) the CMSSM and (right) the NUHM1. 
The dark shaded area at low $m_0$ and high $m_{1/2}$ is
  excluded due 
  to a scalar tau LSP, and the light shaded areas at low $m_{1/2}$ do not
  exhibit electroweak symmetry breaking. The nearly horizontal line at
  $m_{1/2} \approx 160$~GeV in the lower panel 
  has $m_{\tilde \chi_1^\pm} = 103$~GeV, and the area
  below is excluded by LEP searches. Just above this contour at low $m_0$
  in the lower panel is the region that is
  excluded by trilepton searches at the Tevatron.
  Shown in each plot is the best-fit point, indicated by a filled
  circle, and the 
  68 (95)\%~C.L.\ contours from our fit as dark grey/blue (light
  grey/red) overlays~\protect\cite{MC2}.
  Also shown are 5-$\sigma$ discovery contours at the LHC with
  the indicated luminosities and centre-of-mass energies.
} 
\label{fig:planes}
\end{figure*}

\begin{figure*}[htb!]
\resizebox{6.5cm}{!}{\includegraphics{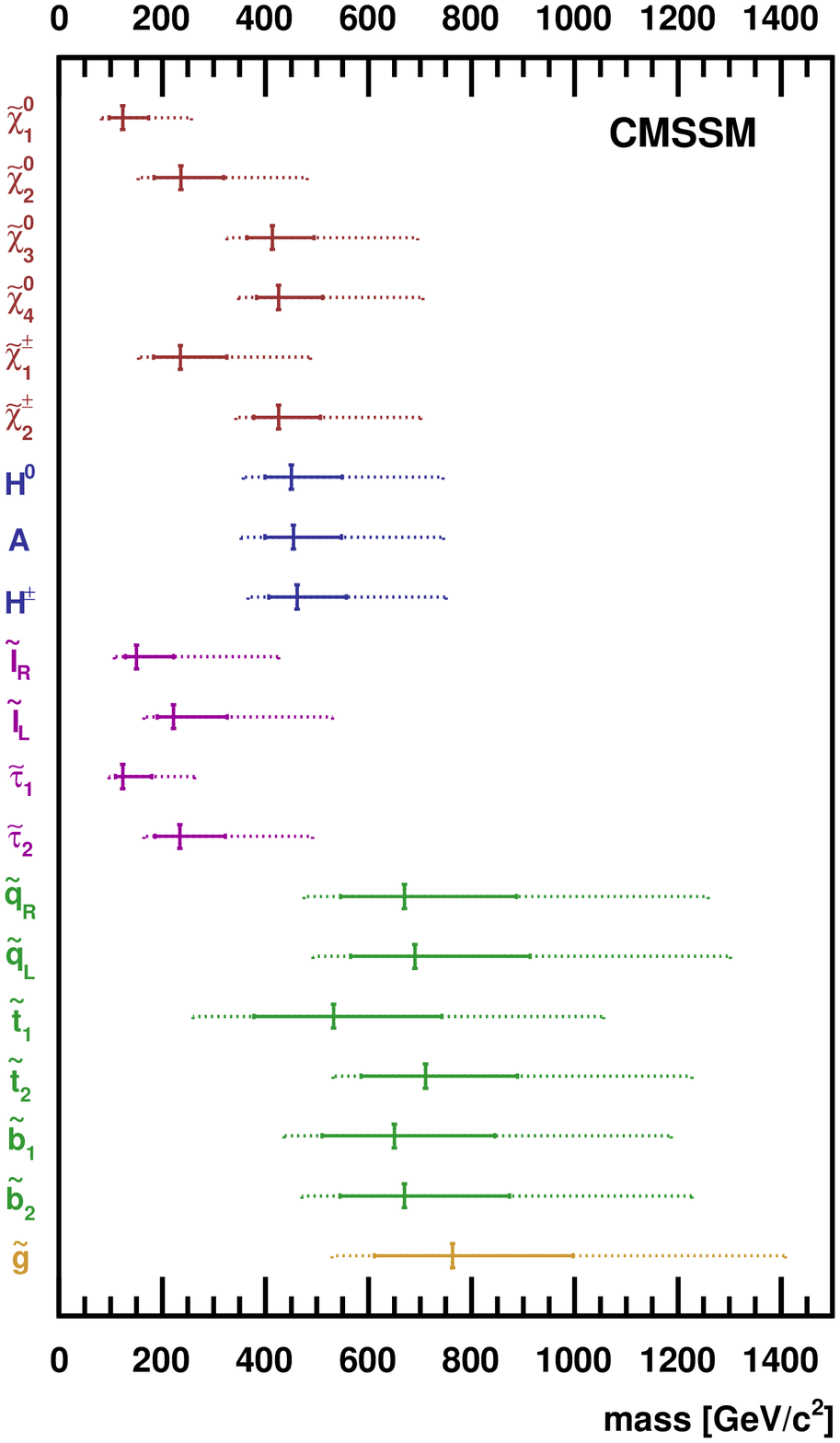}}
\resizebox{6.5cm}{!}{\includegraphics{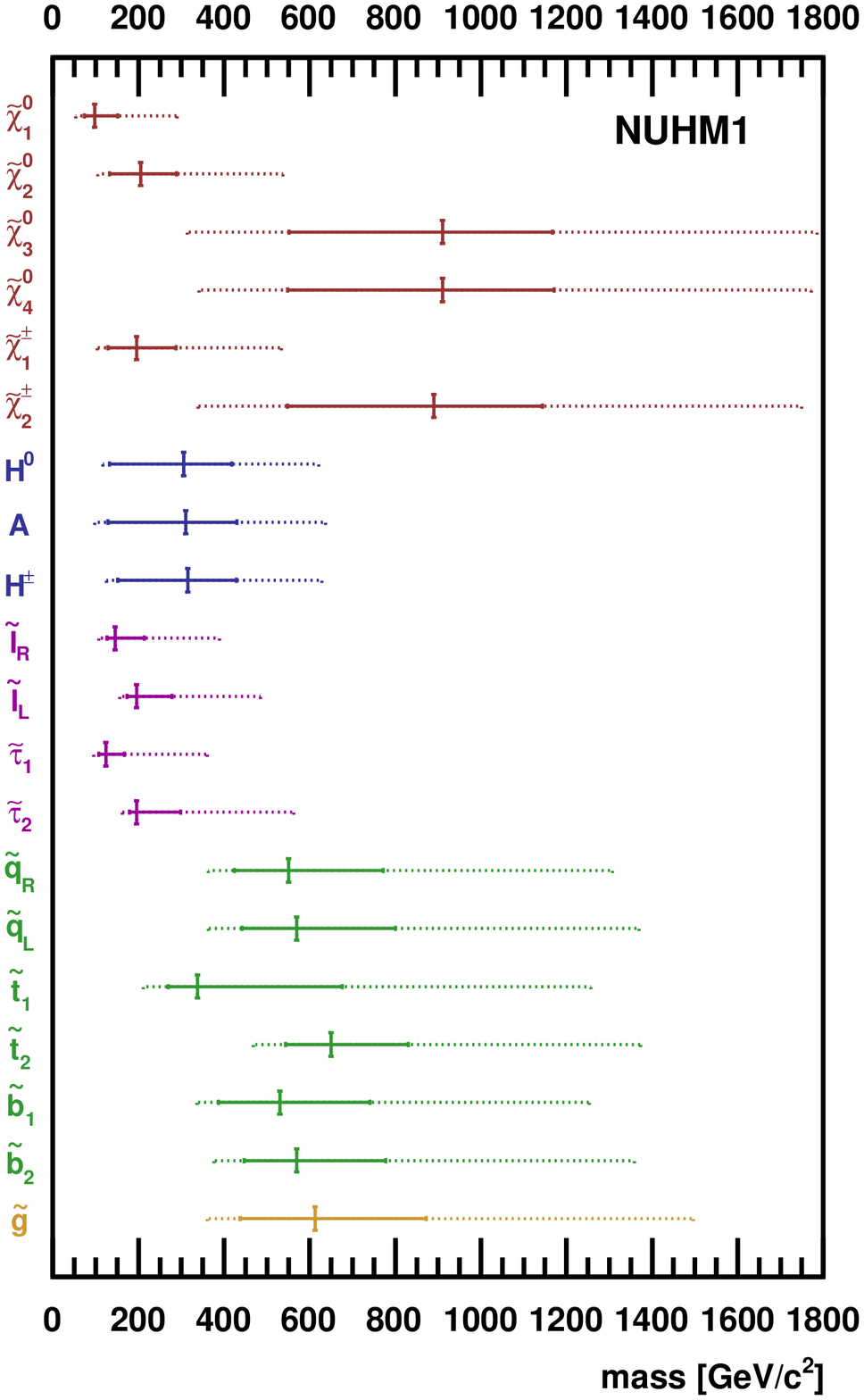}}
\caption{\it Spectra in the CMSSM (left) and the NUHM1 (right). The vertical
solid lines indicate the best-fit values, the horizontal solid lines
are the 68\% C.L.\
ranges, and the horizontal dashed lines are the 95\% C.L.\ ranges for the
indicated mass parameters~\protect\cite{MC3}. 
}
\label{fig:spectra}
\end{figure*}

\begin{figure*}[htb!]
{\resizebox{6.5cm}{!}{\includegraphics{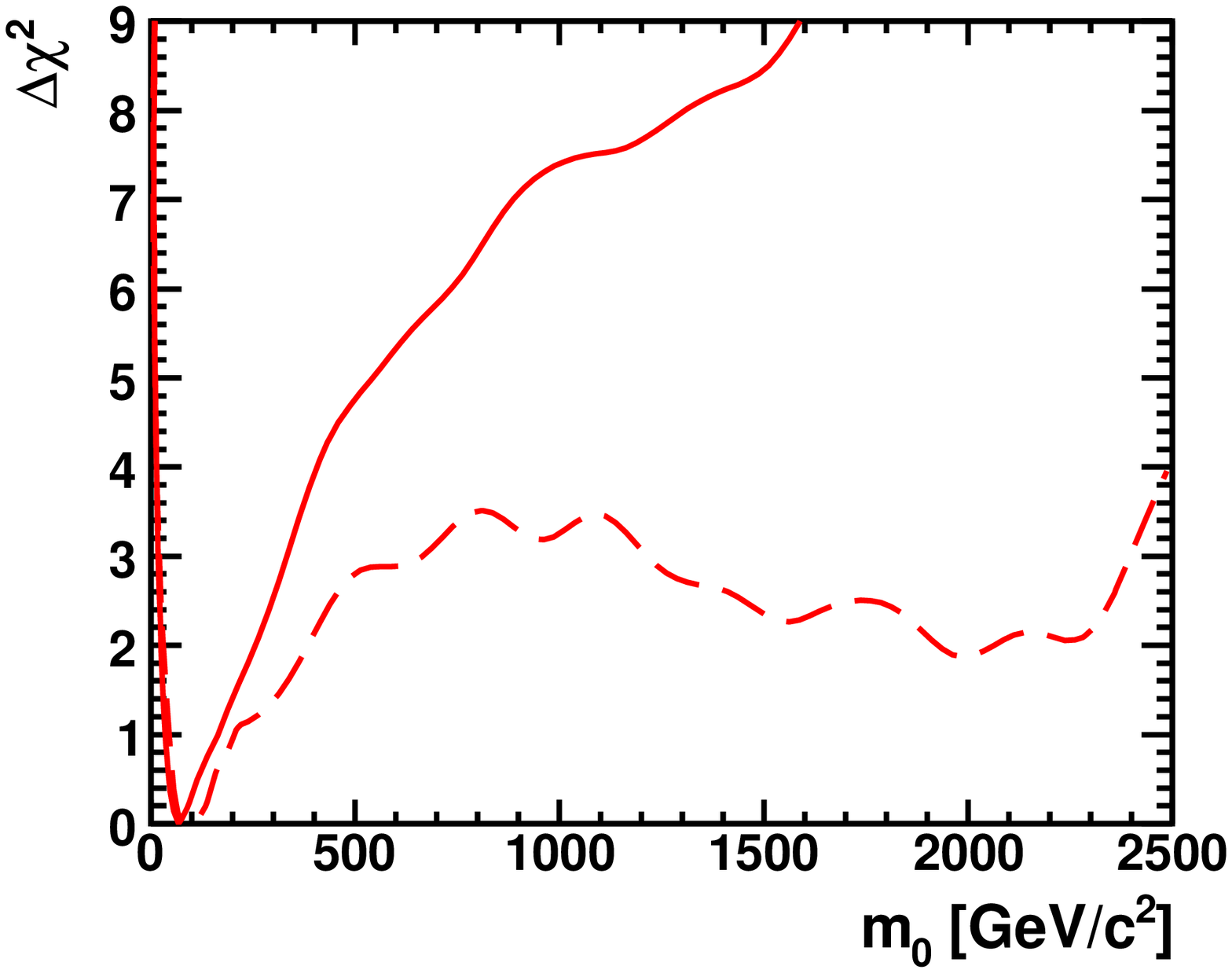}}}  
{\resizebox{6.5cm}{!}{\includegraphics{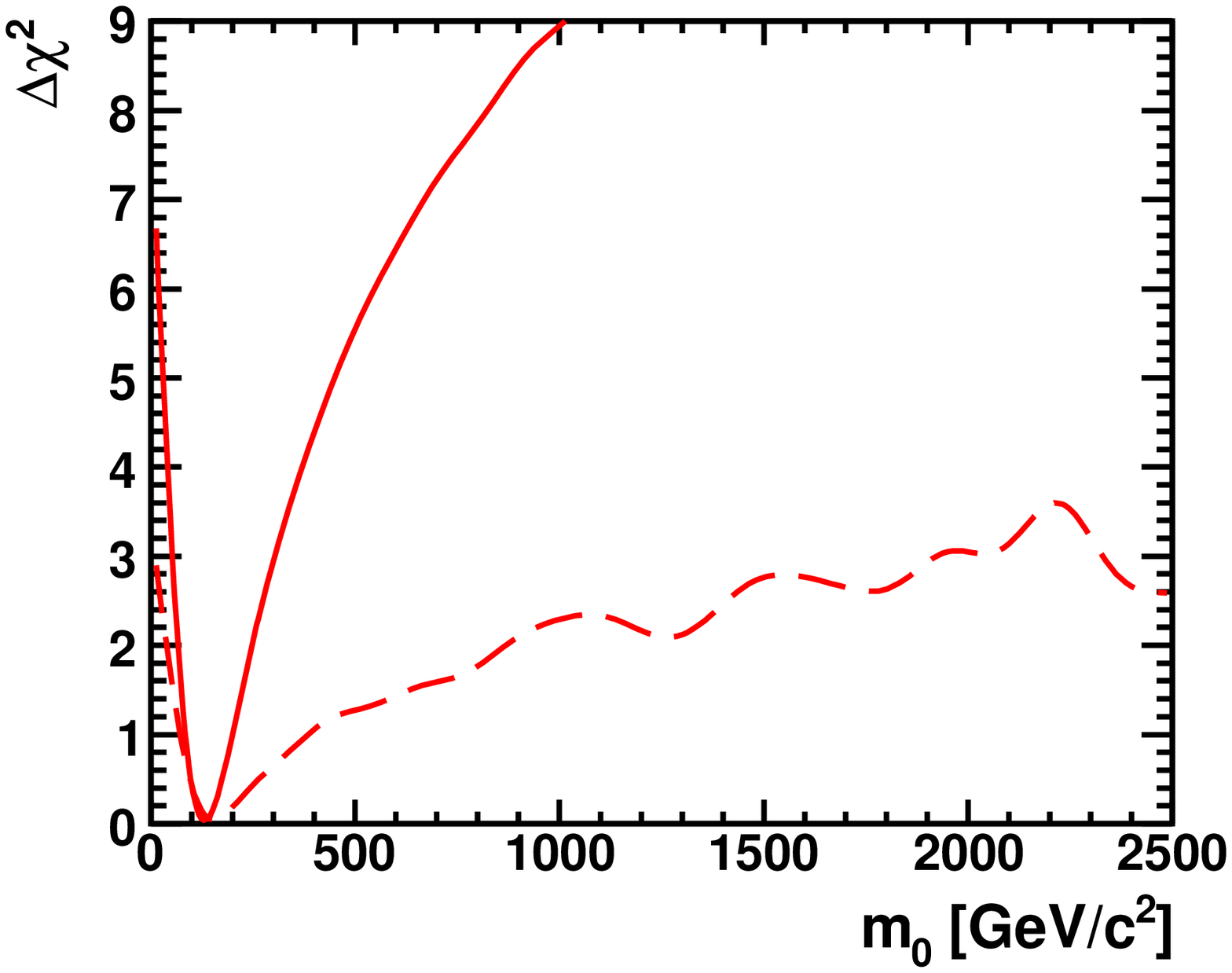}}}
\caption{\it The likelihood functions for $m_0$ in the CMSSM (left plot) and
  in the NUHM1 (right plot). The $\chi^2$ values are shown including
  (excluding) the $g_\mu - 2$ constraint as the solid (dashed) curves~\protect\cite{MC3}.
}
\label{fig:drop}
\end{figure*}

In both the CMSSM and the NUHM1 we found that small $m_{1/2}$ and $m_0$
in the coannihilation region are preferred, with the focus-point region
disfavoured. The best-fit points, 68\% and 95\% CL regions in the $(m_0, m_{1/2})$
planes of the CMSSM and NUHM1 are shown in Fig.~\ref{fig:planes}~\cite{MC2}, and
the corresponding spectra are shown in Fig.~\ref{fig:spectra}~\cite{MC3}. The favoured
areas of the planes shown in Fig.~\ref{fig:planes} are quite sensitive to the
treatments of the constraints, particularly $g_\mu - 2$ and $b \to s \gamma$~\cite{MC2}.
In the extreme case when the $g_\mu - 2$ constraint is dropped entirely, as
in Fig.~\ref{fig:drop}, large values of $m_0$ are no longer strongly
disfavoured, although the other constraints still show some slight preference
for small $m_0$~\cite{MC3}.

Fig.~\ref{fig:planes} also shows the expected sensitivity of the LHC for a
discovery of supersymmetry with 5-$\sigma$ significance for varying LHC
energies and luminosities. We see that there may be a fair chance to discover 
supersymmetry even in early LHC data. However, at the 95\% CL,
supersymmetry might still lie beyond the reach of the LHC with 1/fb of
data at 14~TeV, as could also be inferred from the 95\% CL ranges
in Fig.~\ref{fig:spectra}.

\section{Detecting Supersymmetric Dark Matter}

Several strategies for the detection of WIMP dark matter particles
such as the LSP have been proposed, including the direct search
for scattering on nuclei in the laboratory~\cite{GW}, the search for energetic
neutrinos produced by WIMP annihilations in the core of the Sun or Earth~\cite{nus},
the search for energetic photons produced by WIMP annihilations
in the galactic centre or elsewhere in the Universe~\cite{gammas}, and the searches
for positrons, antiprotons, etc., produced by WIMP annihilations in the
galactic halo~\cite{pbars}.

\begin{figure*}[htb!]
\resizebox{6.5cm}{!}{\includegraphics{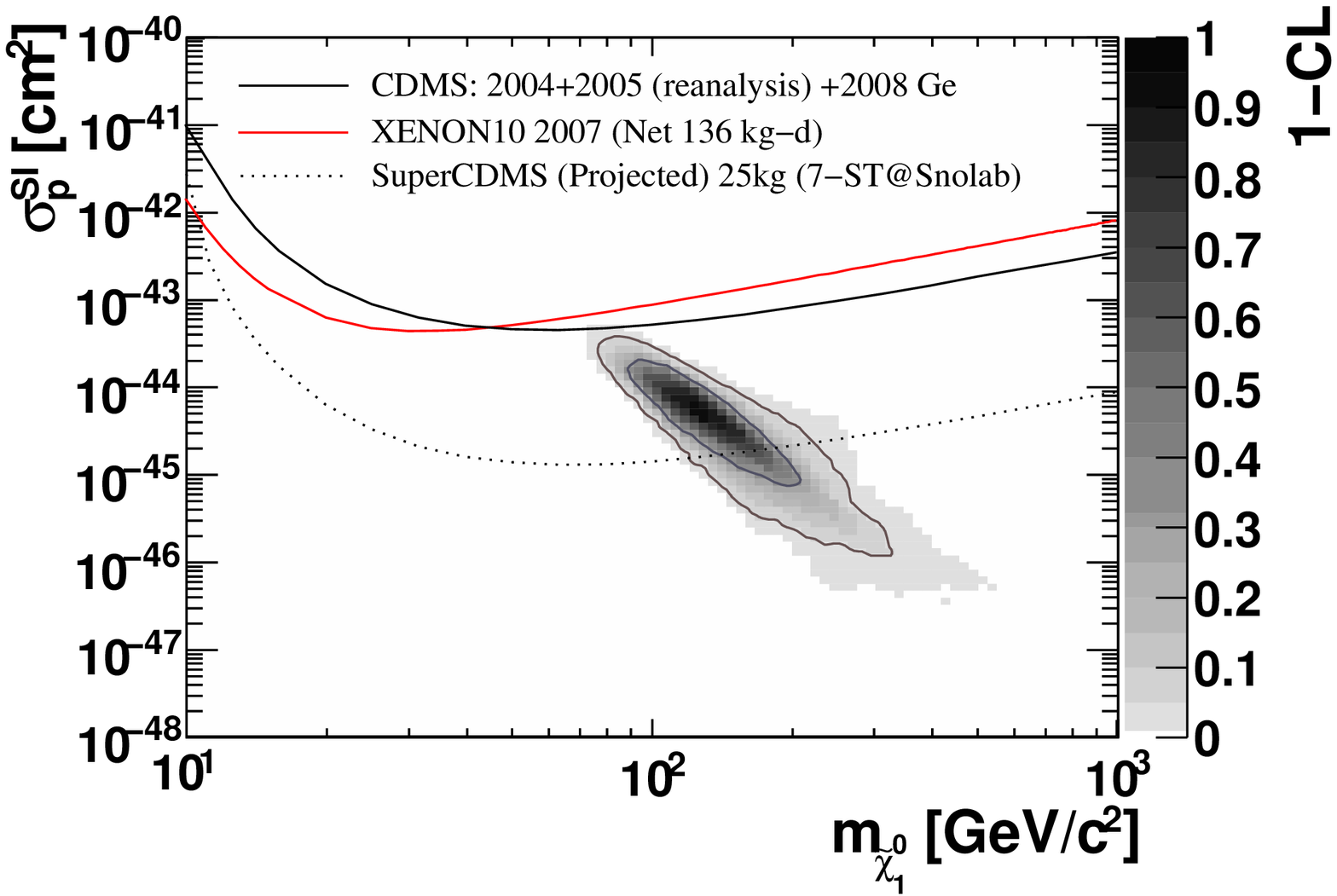}}
\resizebox{6.5cm}{!}{\includegraphics{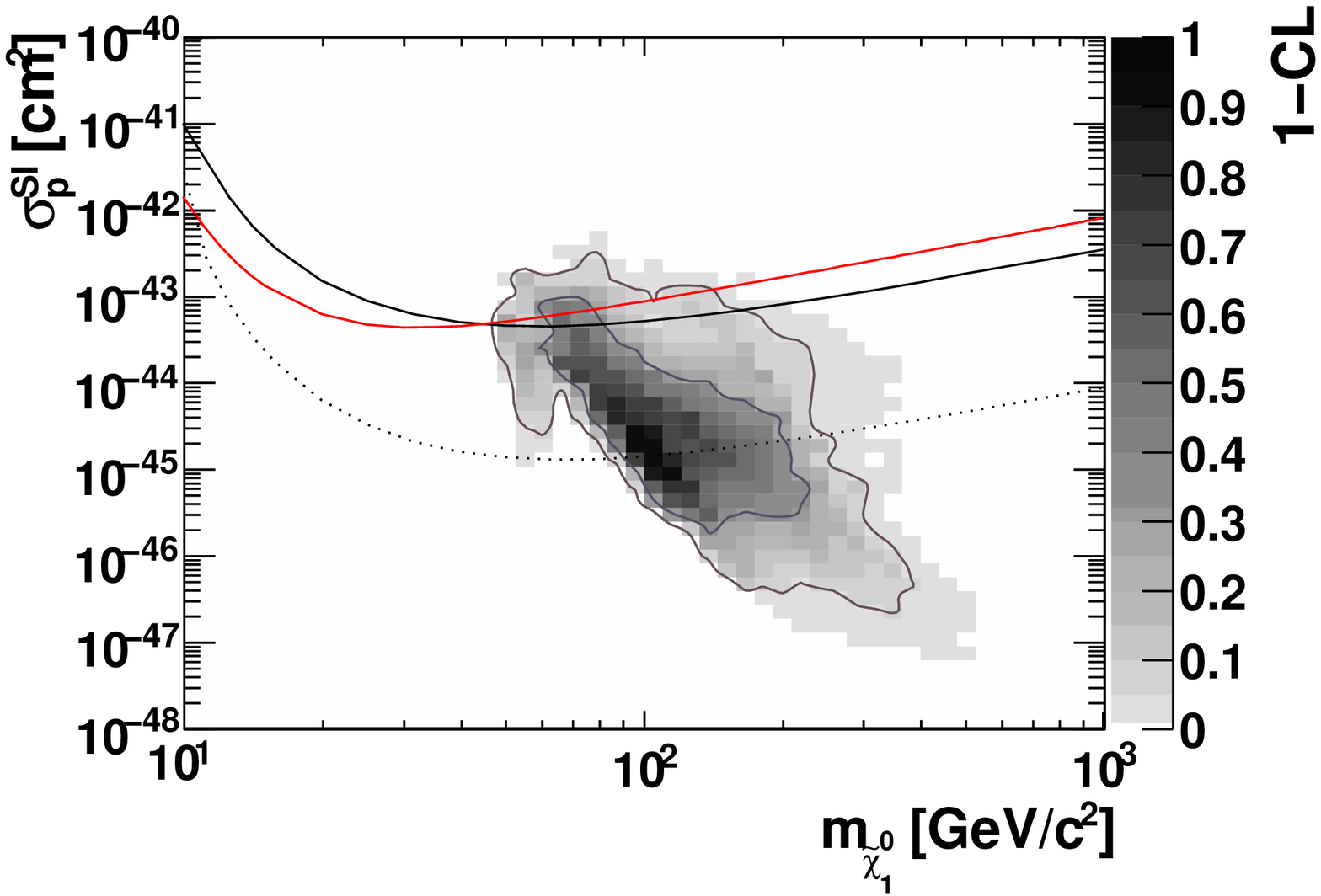}}
\vspace{-1cm}
\caption{\it The correlation between the spin-independent dark matter
scattering cross section and $m_\chi$
in the CMSSM (left panel) and in the NUHM1 (right panel). The solid lines~\protect\cite{DMtool}
are the experimental upper limits from CDMS~\protect\cite{CDMS} and Xenon10\protect\cite{Xe10}, 
The dashed line~\protect\cite{DMtool} indicates the projected sensitivity of the SuperCDMS 
experiment~\protect\cite{superCDMS}: that of Xenon100 may be similar.
}
\label{fig:scattering}
\end{figure*}

As seen in Fig.~\ref{fig:scattering}, within the global fits to
supersymmetric model parameters discussed earlier, our predictions for the direct
nuclear scattering rates in the CMSSM and NUHM1 lie somewhat
below the sensitivities of the CDMS and Xenon10 experiments, 
but within reach of planned upgrades of these experiments~\cite{MC3}. 
Subsequently, the CDMS~II~\cite{CDMSII} and
Xenon100~\cite{Xe100} experiments have announced results with somewhat
improved sensitivity. In particular, the CDMS II experiment reported
two events with relatively low recoil energies (corresponding possibly to
the scattering of a WIMP weighing $< 30$~GeV) where less than one event
was expected~\cite{CDMSII}, but this hint was not confirmed by the Xenon100 experiment
in its initial 11-day test run~\cite{Xe100}. (Nor have possible signals in the DAMA/LIBRA~\cite{DAMA} and 
CoGeNT experiments~\cite{CoGeNT} been confirmed by either CDMS or Xenon100.) It is
expected that updated Xenon100 results with much greater sensitivity will be
announced soon, reaching significantly into the scattering rates expected
within our global fits. (Though it
should be noted that these predictions assume one particular value
for the spin-independent scattering matrix element, which is a significant
source of uncertainty in the predictions~\cite{EOSavage1}.)

\begin{figure*}
\begin{center}
\resizebox{11cm}{!}{\includegraphics{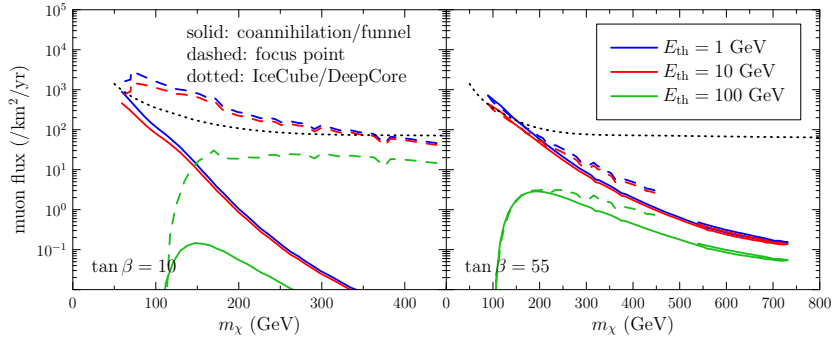}} 
\end{center}
 \caption{\it   
    The CMSSM muon fluxes though a detector calculated for $A_0 = 0$
    and (left) $\tan \beta = 10$, (right) $\tan \beta = 55$,
    along the WMAP strips in the coannihilation/funnel regions (solid)
    and the focus-point region (dashed)~\protect\cite{EOSavage}.
    Fluxes are shown for muon energy thresholds of (top to bottom)
    1~GeV, 10~GeV, and 100~GeV.
     Also shown is a conservative estimate of the sensitivity of the
    IceCube/DeepCore detector (dotted)~\protect\cite{ICDC}, normalized to a muon threshold
    of 1~GeV, for a particular hard annihilation spectrum that is a rough
    approximation to that expected in CMSSM models.
    }
  \label{fig:nuflux}
\end{figure*}

The next most promising strategy for indirect detection of dark matter
may be the search for energetic neutrinos emitted by WIMP annihilations
in the core of the Sun~\cite{nus}. It is often assumed that the annihilation rate is
in equilibrium with the WIMP capture rate, but this is not true in general
in the CMSSM~\cite{EOSavage}. Nor is spin-dependent scattering the dominant
mechanism for LSP capture by the Sun, as is often assumed:
spin-independent scattering on heavier elements inside the Sun may
also be important, even dominant~\cite{EOSavage}. As seen in Fig.~\ref{fig:nuflux},
in a general survey of the CMSSM
parameter space~\cite{EOSavage}, we find significant portions of the focus-point strips,
and some parts of the coannihilation strips, where the flux of energetic 
neutrinos due to LSP annihilations may be detectable in the IceCube/DeepCore
experiment~\cite{ICDC}.

\section{The Start-up of the LHC}

The LHC made its first collisions on November 29th, 2009, and its first 7-TeV
collisions on March 30th, 2010. Much jubilation, but where are the Higgs boson
and supersymmetry, you may ask. It should be recalled that the total
proton-proton cross section for producing a new particle weighing $\sim 1$~TeV
is $\sim 1$/TeV$^2$, possibly suppressed even further by small couplings
$\sim \alpha^2$, whereas the total cross section $\sim 1/m_\pi^2$, so that
the `interesting' new physics signal is likely to occur in $ \sim 10^{12}$ of
the collisions. This is like looking for a needle in $\sim 100,000$ haystacks!

So far the LHC experiments have seen only a few $\times 10^{12}$
collisions. The missing $E_T$ distribution agrees perfectly with simulations over 
more than 6 orders of magnitude~\cite{MET}, and there is no sign yet of an excess of
events that might be due to the production and escape of dark matter
particles, whether they be LSPs, LKPs, LTPs, or whatever. Moreover,
the kinematics of the events with missing $E_T$ is exactly what one
would expect from mismeasured QCD events and other Standard Model
backgrounds: no signs yet of new physics beyond the Standard Model.

\begin{figure*}[htb!]
\begin{center}
\resizebox{8cm}{!}{\includegraphics{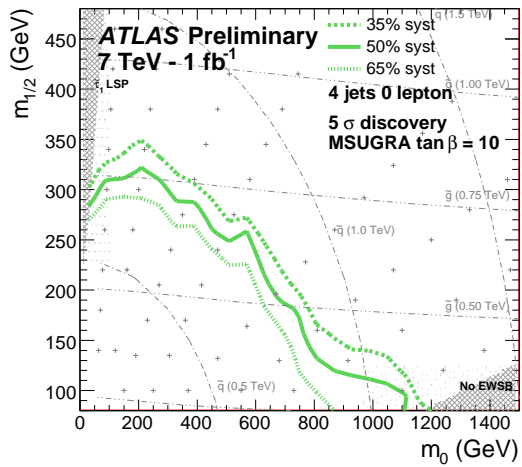}}
\end{center}
\caption{\it The sensitivity in the $(m_0, m_{1/2})$ plane
of the ATLAS experiment for a 5-$\sigma$ discovery of a
supersymmetric signal with 1/fb at 7 TeV in the centre of mass.
}
\label{fig:7TeV}
\end{figure*}

The results of our frequentist likelihood analysis were compared in
Fig.~\ref{fig:planes} with the estimated sensitivity of the LHC at or
close to its design energy. In 2010 it has been operating at $\sim 7$
TeV and accumulating $\sim 50$/pb of integrated luminosity, which is
sufficient to extend the reach for supersymmetry beyond the Tevatron.
The centre-of-mass energy may be increased in 2011 to 8 or 9 TeV,
and the objective is to accumulate $\sim 1$/fb of integrated
luminosity. Fig.~\ref{fig:7TeV} shows the estimated sensitivity of 
supersymmetry searches with the ATLAS experiment~\cite{ATLAS7} using 1/fb of data at 7 TeV.
Comparing with Fig.~\ref{fig:planes}, we see that the best-fit points in the
CMSSM and NUHM1 should lie within reach. There are significant prospects
for soon getting some interesting news about supersymmetry, one way or
the other.

\section{A Conversation with Mrs. Thatcher}

In 1982, Mrs. Thatcher, the British Prime Minister at the time, visited CERN,
and I was introduced to her as a theoretical physicist. ``What exactly do you do?",
she asked in her inimitably intimidating manner. ``I think of things for experimentalists to look
for, and then I hope they find something different", I responded. Somewhat
predictably, Mrs. Thatcher asked ``Wouldn't it be better if they found what
you predicted?" My response was that "If they found exactly what the theorists
predicted, we would not be learning so much". In much the same spirit, I hope
(and indeed expect) that the LHC will become most famous for discovering
something that I did NOT discuss in this talk - as long as it casts light on dark matter!

\end{document}